\documentclass[preprint,aps,prb,floatfix,showpacs]{revtex4}
\usepackage{graphicx}
\usepackage{dcolumn}
\usepackage{bm}
\usepackage{amsmath}
\usepackage{subfigure}

\newcommand{\bM}{\mathbf{M}}
\newcommand{\bv}{\mathbf{v}}
\newcommand{\bs}{\mathbf{s}}
\newcommand{\bc}{\mathbf{c}}
\newcommand{\bsh}{\hat{\mathbf{S}}}
\newcommand{\bzh}{\hat{\mathbf{z}}}

\newcommand{\bMh}{\hat{\mathbf{M}}}
\newcommand{\bnh}{\hat{\mathbf{n}}}
\newcommand{\br}{\hat{\mathbf{r}}}
\newcommand{\sh}{\hat{S}}
\newcommand{\Mh}{\hat{M}}
\newcommand{\RR}{\mathcal{R}}
\newcommand{\HH}{\mathcal{H}}
\newcommand{\bT}{\mathbf{T}}
\newcommand{\norm}[1]{\lVert#1\rVert}
\newcommand{\abs}[1]{\lvert#1\rvert}
\DeclareMathOperator{\sign}{sign}

\DeclareMathOperator{\Ln}{Ln}
\newtheorem{lemma}{Lemma}

\begin{document}

\preprint{APS/123-QED}

\title{
	Constrained Monte Carlo Method and \\
    Calculation of the Temperature Dependence of Magnetic Anisotropy
}

\author{P. Asselin}
\affiliation{Seagate Technology, Bloomington, MN 55435, USA}
\author{R. F. L. Evans}
\email{rfle500@york.ac.uk}
\author{J. Barker}
\author{R. W. Chantrell}
\affiliation{Department of Physics, University of York,	Heslington, York YO10 5DD United Kingdom}
\author{R. Yanes}
\author{O. Chubykalo-Fesenko}
\affiliation{Instituto de Ciencia de Materiales de Madrid, CSIC, Cantoblanco, Madrid 28049, Spain}
\author{D. Hinzke}
\author{U. Nowak}
\affiliation{Universitat Konstanz, Fachbereich Physik, Universit\"atsstra{\ss}e 10, D-78464 Konstanz, Germany}

\date{\today}

\begin{abstract}
We introduce a constrained Monte Carlo method which allows us to traverse the phase space of a classical spin system while fixing the magnetization direction. Subsequently we show the method's capability to model the temperature dependence of magnetic anisotropy, and for bulk uniaxial and cubic anisotropies we recover the low-temperature Callen-Callen power laws in $M$. We also calculate the temperature scaling of the 2-ion anisotropy in L$1_0$ FePt, and recover the experimentally observed $M^{2.1}$ scaling. The method is newly applied to evaluate the temperature-dependent effective anisotropy in the presence of the N\'eel surface anisotropy in thin films with different easy axis configurations. In systems having different surface and bulk easy axes, we show the capability to model the temperature-induced reorientation transition. The intrinsic surface anisotropy is found to follow a linear temperature behavior in a large range of temperatures.
\end{abstract}

\maketitle

\section{Introduction}
The temperature dependence of magnetocrystalline anisotropy of pure ferromagnets has been well known for decades, following the Callen-Callen theory \cite{CallenCallen}. Examples of such systems classed as pure ferromagnets in the anisotropic sense include Gadolinium (Uniaxial) and Fe (Cubic). Other ferromagnets, such as Co, have a much more complicated temperature dependence, due to the crystallographic origin of the anisotropy \cite{Bruno1993}. Indeed, with increased temperature the anisotropy in Co exhibits a change in sign, indicating a transformation from an easy-axis to easy-plane anisotropy. Such behavior is not explained by the Callen-Callen theory. Other materials not exhibiting a simple temperature dependence of the anisotropy are magnetic transition metal alloys such as FePt and CoPt. Here the origin of the anisotropy is due to an ion-ion anisotropic exchange interaction, arising from the underlying crystal symmetry. In FePt the anisotropy exhibits an unusual temperature dependence\cite{Thiele,Okamoto} of $K_{\mathrm{u}}^{\mathrm{eff}} \propto M^{2.1}$. Other systems of technological interest include magnetic thin films and nanoparticles, where surface effects can lead to unusual temperature dependent anisotropies. In general the temperature dependence of anisotropy in many materials is not obvious, and as such is still an active area of research some 40 years after the Callen-Callen theory.

Recently, the high temperature behavior of magnetic anisotropy has become important due to the applications in heat-assisted magnetic recording (HAMR) \cite{HAMR1,HAMR2,HAMR3}. The idea of HAMR is based on the heating of the recording media to decrease the writing field of the high anisotropy media (such as FePt) to values compatible with the writing fields provided by conventional recording heads. Since the writing field is proportional to the anisotropy field $H_{\mathrm{k}}=2K_{\mathrm{u}}^{\mathrm{eff}}(T)/M(T)$, the knowledge of the scaling behavior of the anisotropy $K_{\mathrm{u}}$ with the magnetization $M$ has become a paramount consideration for HAMR\cite{Lyberatos}. It should be noted that even in relatively simple systems, a simple scaling behavior predicted by the Callen-Callen theory is only valid at temperatures far from the Curie temperature. The systems proposed for HAMR applications can also include more complex composite media such as soft/hard bilayers \cite{Sanchez}, FePt/FeRh with metamagnetic phase transition \cite{Thiele2003,Guslienko}, or exchange-bias systems \cite{EvansEPL2009}.

The evaluation of the temperature dependence of magnetic anisotropy is also important for the modeling of the laser-induced demagnetization processes. The thermal decrease of the anisotropy during the laser-induced demagnetization has been shown to be responsible for the optically-induced magnetization precession \cite{Atxitia2007}. Thus the ability to evaluate the temperature dependence of the anisotropy in complex systems at arbitrary temperatures is highly desired from the fundamental and applied perspectives.

In this sense magnetic thin films with surface anisotropy are a representative example of this more complicated situation. In ultra-thin films, especially when in contact with a different non-magnetic matrix, the interplay between the broken symmetry, magnetostriction, roughness, spin-orbit interaction and charge transfer can often be encompassed in a phenomenological model as an additional surface anisotropy. Since the surface anisotropy has a different temperature dependence from the bulk, multiple experiments on thin film have demonstrated the occurrence of the spin reorientation transition from an out-of-plane to in-plane magnetization both as a function of temperature and thin film thickness \cite{Schultz, Berger, Baek2003, Bruno1989, Allenspach1992, Enders, Dinia1998,Farle1998}. The possibility to engineer the reorientation transition also requires the capability to evaluate the temperature dependence of the surface anisotropy independently from the bulk.

In the following we present a new Monte Carlo method which can be applied to the computation of both bulk and surface anisotropies at finite temperature. This article represents a first step showing the possibility to calculate the temperature-dependent anisotropies in principle.  When combined with detailed magnetic information, such as that available from ab-initio methods\cite{KazantsevaPRB2008}, this forms a very powerful method of engineering the temperature dependent properties of a magnetic system.

\section{Modeling Methods}\label{sec:method}
For the calculations presented in the following we describe the magnetic properties of the system by utilizing a classical atomistic spin model, similar to Nowak \cite{NowakReview}, with a general Hamiltonian of the form:
\begin{eqnarray}\label{eq:ham}
  \nonumber \mathcal{H} &=& - \sum_{i \ne j} J_{ij} \bsh_i \cdot \bsh_j - K_{\mathrm{u}} \sum_{i} {\sh_{iz}}^2 \\
  \nonumber            && - \frac{K_{\mathrm{c}}}{2} \sum_{i} \left({\sh_{ix}}^4 +  {\sh_{iy}}^4 + {\sh_{iz}}^4\right) \\
               && - \sum_{i \ne j} \frac{K_{\mathrm{s}}}{2} \left(\bsh_i \cdot \br_{ij}\right)^2
\end{eqnarray}
describing the exchange interaction ($J_{ij}$), uniaxial ($K_{\mathrm{u}}$), cubic ($K_{\mathrm{c}}$), and N\'eel surface anisotropies \cite{Neel1954} ($K_{\mathrm{s}}$) respectively. Note that in the following text, we also refer to the effective temperature dependent values $K_{\mathrm{u}}^{\mathrm{eff}}$ (uniaxial), $K_{\mathrm{c}}^{\mathrm{eff}}$ (cubic), and $K_{\mathrm{s}}^{\mathrm{eff}}$ (surface). The summations for the exchange and surface anisotropy are generally limited to nearest neighbors only, except for the case of FePt where the full exchange up to 5 neighboring cells (around 1300 neighbors) was taken into account.

The parameters are chosen to represent a generic ferromagnet, with a Curie temperature($T_{\mathrm{c}}$) of around 1000K, and arbitrary anisotropy constants, where $K_{\mathrm{u}},K_{\mathrm{c}} \ll J_{ij}$. Note that all three anisotropy terms have been included within the same Hamiltonian for brevity -- in practice a system will only have cubic or uniaxial anisotropy, and clearly only surface atoms will possess surface anisotropy.

We compute thermodynamic properties by averaging over the Boltzmann distribution using the Metropolis algorithm\cite{Metropolis1953}. Our innovation, which we call the constrained Monte Carlo method (CMC), is to modify the elementary moves of the random walk so as to conserve the average magnetization direction $\bMh\equiv\bigl(\sum_i\bsh_i\bigr)/\norm{\sum_i\bsh_i}$. In this way we sample the Boltzmann distribution over a submanifold of the full phase space. Thus we keep the system out of thermodynamic equilibrium in a controlled manner, while allowing its microscopic degrees of freedom to thermalize.

Because the system cannot reach full equilibrium, the average of the total internal torque $\bT = \bigl\langle-\sum_i\bsh_i\times\partial \mathcal{H}/\partial\bsh_i\bigr\rangle$ does not vanish. We show in Appendix \ref{cmc:torques} that this is equal to the macroscopic torque $-\bMh\times\partial\mathcal{F}/\partial\bMh$, where $\mathcal{F}(\bMh)$ is the Helmholtz free energy, now a function of $\bMh$. Even though we cannot compute $\mathcal{F}$ directly, we can reconstruct its angular dependence by integration,

 \begin{equation}\label{eqn:TfromF}
    \mathcal{F}(\bMh) = \mathcal{F}(\bMh_0) + \int_{\bMh_0}^{\bMh} \left(\bMh' \times \bT' \right) \cdot d\bMh'
 \end{equation}

where the integral on $\bMh'$ can be taken along any path on which the system behaves reversibly. This in turn gives us the anisotropy constants at any temperature.

In practice it is often simpler to recover the anisotropy constants directly from the derivatives. We first initialize the system with uniform magnetization in a direction of our choice, away from the anisotropy axes, where we expect a nonzero torque. Next we evolve the system by constrained Monte Carlo until the length of the magnetization reaches equilibrium. We then take a thermodynamic average of the torque over a large number constrained Monte Carlo steps, typically 50,000. We repeat at other orientations and we finally reconstruct the anisotropy constants from the angular dependence of the torque.

\section{Constrained Monte Carlo}\label{sec:CMC}
The Metropolis algorithm works by generating trial moves at random and accepting or rejecting each move based on the ratio of the Boltzmann probability densities $\exp(-\beta \HH)$, $\beta=1/kT,$ at the initial and final states.
This ratio depends only on the energy difference between the two states. An accepted move yields a new state and a rejected move yields a repetition of the initial state (``null move'' in the following). There is considerable freedom in the construction of trial moves. It is required only that each move have the same probability density
as the inverse move (reversibility), and that all states be reachable by a sequence of moves (ergodicity). Under these conditions the random walk's limiting distribution is the Boltzmann distribution.

In the usual Monte Carlo method we generate our trial moves by drawing a vector $\bv$ from an isotropic normal distribution, choosing a spin $\bsh_i$ at random, adding $\bsh_i\times\bv$ to it and normalizing the result
to obtain a trial spin $\bsh_i'$. The probability density of the move depends only on the angle between $\bsh_i$ and $\bsh_i'$, which ensures reversibility. Ergodicity is obvious. The variance of the $\bv$ distribution controls the size of the attempted moves and can be chosen at will to improve the ratio of accepted to rejected moves, similarly to the parameter $\alpha$ in reference \onlinecite{Metropolis1953}. For our Hamiltonian, the energy difference involves only spin $i$ and the few neighboring spins to which it is coupled by exchange, so the decision to accept or reject the move can be made quickly. A sequence of $N$ moves, counting null moves, constitutes a {\em step}; we compute quantities of interest once per step to average them.

In the constrained Monte Carlo method the trial moves act on two spins at a time. The extra degrees of freedom allow us to fix $\bMh$ to any given unit vector, which we take here to be the positive $z$ axis since we can always reduce the problem to this case by means of a global rotation. Ignoring the $z$ coordinates for a moment, we simply displace two spins by equal and opposite amounts in the $XY$ plane. There are technical issues due to the fact that $\sh_x,\sh_y$ are not canonical coordinates; furthermore, we need to allow sign changes in the $z$ coordinates.
In the end we settled on the following:

\begin{enumerate}\setlength{\itemsep}{0pt}
    \item Choose a {\em primary spin} $\bsh_i$ and a
	{\em compensation spin} $\bsh_j,$ not necessarily neighbors.
    \item Displace the primary spin as in the usual Monte Carlo
	method, obtaining a new spin $\bsh_i'$.
	\label{alg:primary}
    \item Adjust the compensation spin's $x$ and $y$ components
	to preserve $M_x=M_y=0$,
    \label{alg:compensate}
    \begin{flalign*}
	\sh_{jx}' &= \sh_{jx} + \sh_{ix} - \sh_{ix}'\\
	\sh_{jy}' &= \sh_{jy} + \sh_{iy} - \sh_{iy}'
    \end{flalign*}
    \item Adjust the $z$ component, \label{alg:adjustz}
    \begin{equation*}
	\sh_{jz}' =
	\sign\left(\sh_{jz}\right)\sqrt{1-\sh_{jx}'^2-\sh_{jy}'^2}
	\quad.
    \end{equation*}
    If the argument of the square root is negative, stop and take a null move.
    \item Compute the new magnetization, \label{alg:Mz}
	\begin{equation*}
	    M_z'= M_z + \sh_{iz}' + \sh_{jz}' - \sh_{iz} - \sh_{jz} \quad.
	\end{equation*}
	If $M_z'\leq 0$, stop and take a null move.
    \item Compute the energy difference $\Delta \HH=\HH'-\HH$.\label{alg:dE}
    \item Compute the acceptance probability $P$, \label{alg:Boltzmann}
    \begin{equation*}
	P=\min \Biggl[1,
	    \biggl(\frac{M_z'}{M_z}\biggr)^2
	    \frac{\abs{\sh_{jz}}}{\abs{\sh_{jz}'}}
	    \exp\bigl(-\beta\Delta \HH\bigr)
	\Biggr] \quad.
    \end{equation*}
    \item \label{alg:accept}
    Accept the move with probability $P$
    or take a null move with probability $1-P$.
\end{enumerate}

In effect, we use the compensation spin to project the system back to its admissible manifold. The projection is not orthogonal and does not preserve measure. Consequently the Boltzmann ratio in step \ref{alg:Boltzmann} is multiplied by a geometric correction, the ratio of two Jacobians, which we derive in Appendix \ref{sec:CMCderiv}. We prove ergodicity in Appendix \ref{sec:ergodic}.

The null moves at step \ref{alg:adjustz} handle the kinematic constraints in a natural way. We could instead add fictitious states that allow step \ref{alg:adjustz} to complete and assign zero probability (infinite energy) to these states; this would guarantee rejection at step \ref{alg:accept} and it is simpler to stop at step \ref{alg:adjustz}. Similarly at step \ref{alg:Mz} we reject trial move that would change the sign of $\bMh$; here the end states exist but we simply want to sample them with zero probability.

\section{Calculation of Bulk Anisotropies}\label{sec:tests}
Given the originality of the constrained Monte Carlo method, it is important to ensure that the method is reliable and conforms to existing results, especially regarding the low temperature dependence of bulk anisotropy as predicted by Callen and Callen \cite{CallenCallen}, where uniaxial anisotropy was shown to have an $M^3$ dependence, and cubic anisotropy to have an $M^{10}$ dependence.

\begin{figure}[htpbc]
   \begin{center}
   \includegraphics[totalheight=0.25\textheight]{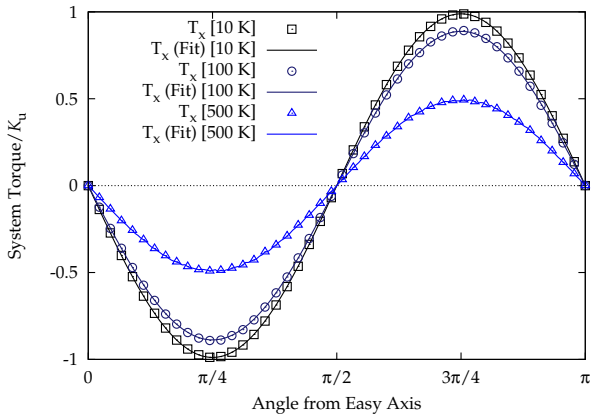}\\
   \caption{
	Simulated and analytical angular dependence of the
	restoring torque for uniaxial anisotropy at temperatures
	10 K, 100 K, and 500 K. (Color Online)
   }
   \label{fig:torque}
   \end{center}
\end{figure}

In the present work such bulk systems were approximated by simulating a generic ferromagnetic system with 16000 spin moments with periodic boundary conditions, to eliminate surface effects and minimize finite size effects. Sample torque curves for the system with uniaxial anisotropy are shown in Fig.~\ref{fig:torque}. The points show the calculated torque and the curves are fits to a $\sin\left(2\theta\right)$ dependence, where $\theta$ is the angle from the easy axis.
The $\sin\left(2\theta\right)$ relationship is seen to hold at all temperatures and the fitted proportionality constant gives $K_{\mathrm{u}}^{\mathrm{eff}}(T)$. In a situation such as this, where all the torque curves have the same shape, the anisotropy is described by a single parameter and it is sufficient to compute the torque at the maximum.

In more general cases, such as those with surface anisotropy, it is necessary to compute the torque at several angular positions. Finally, with every new system it is prudent to verify the shape of the torque curves over many angles, both polar and azimuthal, before reducing the number of points to the minimum necessary.

\begin{figure}[htpbc]
	\centering
	\subfigure[ Temperature dependence of magnetization, uniaxial,
		and cubic anisotropies. The lines provide a guide to the eye.]{
		\label{fig:testsa}
		\includegraphics[totalheight=0.25\textheight]{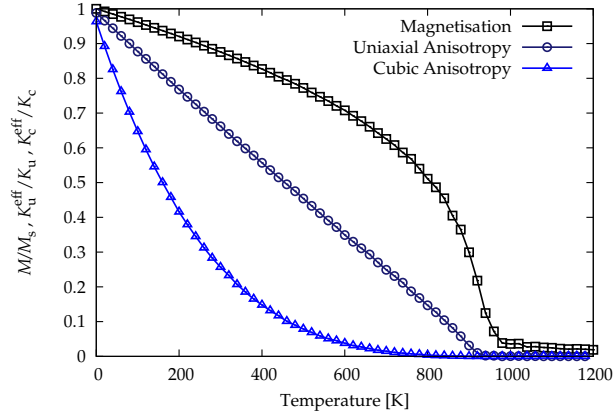}
	}
	\vspace{.0in}
	\subfigure[ Temperature scaling of uniaxial and cubic anisotropies.]{
		\label{fig:testsb}
		\includegraphics[totalheight=0.25\textheight]{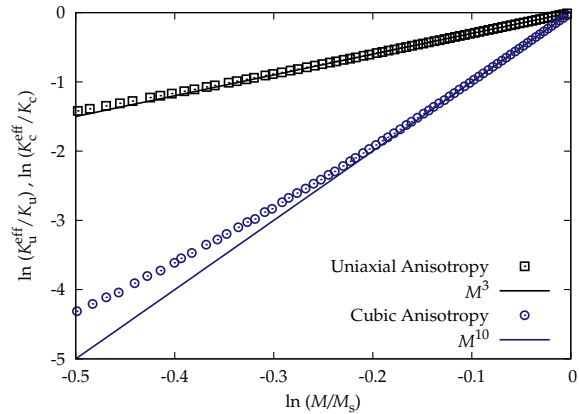}
	}
	\caption[]{Simulation results for the temperature dependence and scaling of a pure ferromagnet with uniaxial
		and cubic anisotropies. (Color Online)}
	\label{fig:unicubic}
\end{figure}

The uniaxial anisotropy, cubic anisotropy and magnetization for the generic system are plotted against temperature in Fig.~\ref{fig:testsa}. In order to reduce the computational effort, the torques were computed at a single angular position, $\theta=45^\circ$ in the uniaxial case and $\theta=22.5^\circ$ in the $xz$-plane for cubic anisotropy. In Fig.~\ref{fig:testsb} the anisotropies are plotted against the magnetization on logarithmic scales.
As can be seen, the results show excellent agreement at low-temperatures with the scaling behavior predicted by Callen and Callen. We note that in these calculations the internal energy and free energy are almost interchangeable. That is, $\partial\mathcal{F}/\partial\theta$ is nearly equal to $\partial\langle \HH\rangle/\partial\theta$. The anisotropy is too weak to affect the entropy and the difference $\langle \HH\rangle-\mathcal{F}=TS$ is nearly independent of $\theta$. We expect this to hold in most magnetic systems, but in full generality the distinction between $\mathcal{F}$ and $\langle \HH\rangle$ must be kept.

In strong anisotropy systems and temperatures close to $T_c$ we have found that the system torque deviates from the expected $\sin (2 \theta)$ behavior. This is in agreement with previous publications where it has been shown that at high temperatures strong magnetization fluctuations lead to several interesting effects, related to the free energy behavior, such as the change of the magnetization length on the saddle point\cite{KazantsevaEPL2009} or the elliptic character of  domain walls\cite{Hinzke2008}.

\section{Calculation of Anisotropy in F\lowercase{e}P\lowercase{t}}\label{sec:FePt}
FePt is a material of current interest because of its extremely high magnetocrystalline anisotropy energy in the L$1_0$ crystal phase\cite{Weller}. The material is also unusual due to the $M^{2.1}$ low-temperature scaling of the effective anisotropy\cite{Thiele,Okamoto}. Ab-initio and Langevin dynamics simulations by Mryasov \emph{et al}\cite{Mryasov} managed to reproduce the observed scaling of the anisotropy by calculation of the internal energy, and in this work we have applied the Constrained Monte Carlo method to the same problem. We have utilized the same Hamiltonian as Mryasov \emph{et al}\cite{Mryasov} and simulated a system 6 nm$^3$ with periodic boundary conditions. The key addition to the generic Hamiltonian in Eq.~(\ref{eq:ham}) is a two-ion anisotropy term of the form:
\begin{equation}\label{eq:2ion}
   \mathcal{H}^{\mathrm{2-ion}} = \sum_{i\ne j} K_{ij}^{(2)} \sh_i^z \sh_j^z
\end{equation}
where $K_{ij}^{(2)}$ is the anisotropy constant which is site and range dependent, as extracted form the ab-initio calculations. The system also possesses an easy-plane anisotropy which is an order of magnitude weaker than the easy-axis 2-ion anisotropy. The existence of these competing anisotropies in fact gives rise to the unusual scaling exponent due to the different temperature scaling of the anisotropic contributions, $M^2$ for two-ion and $M^3$ for single ion.
\begin{figure}[htpbc]
   \begin{center}
   \includegraphics[totalheight=0.25\textheight]{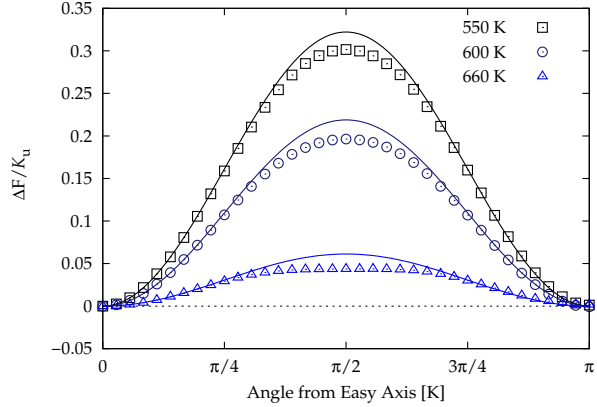}\\
   \caption{Angular dependence of the free energy in L$1_0$ FePt at high temperatures. The curves are $\sin^2\theta$ fits to the data in the range 0:$\pi/5$. (Color Online)}
   \label{fig:FePtTQ}
   \end{center}
\end{figure}
 Due to the large value of the effective anisotropy in FePt, the torque deviates from the expected $\sin(2\theta)$ relationship at elevated temperatures, and so the free energy was obtained by integration over $\theta$. The torque curves were calculated by first equilibrating the system for 10000 steps at each temperature and angle, and then the thermal average of the torque was calculated over a further 70000 steps. Plots of the change in free energy, $\Delta F$, are shown in Fig.~\ref{fig:FePtTQ}, showing a deviation from the usual $\sin^2\theta$ angular dependence at temperatures close to $T_\mathrm{c}$ (700 K). In fact a consistent flattening of the free energy is seen in the proximity of the hard axis - the high anisotropy energy causes a reduction in the length of the magnetization which effectively means it is competing with the exchange interaction. The balance of these effects is that the length of magnetization is slightly lower and therefore so is the torque, leading to a flattening of the free energy in the maximum. This leads to a lower than expected energy barrier which is important for the calculation of relaxation times at high temperature.
\begin{figure}[htpbc]
   \begin{center}
   \includegraphics[totalheight=0.25\textheight]{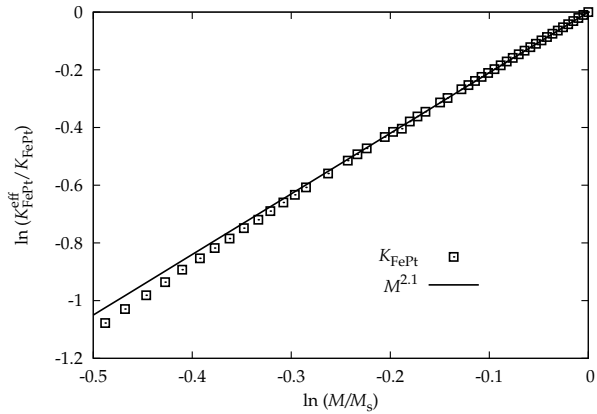}\\
   \caption{Temperature scaling of anisotropy in L$1_0$ FePt.}
   \label{fig:FePt}
   \end{center}
\end{figure}
The low temperature scaling of the anisotropic free energy is plotted in Fig.~\ref{fig:FePt}, showing excellent agreement with the previous theoretical and experimental results\cite{Thiele,Okamoto,Mryasov}.

\section{Thin film systems with N\'eel surface anisotropy}
So far we have demonstrated the ability of the constrained Monte Carlo method to reproduce results which are well known. In the following the method is applied to a system where the temperature dependent behavior is unknown - namely thin films with surface anisotropy. Understanding surface anisotropy presents a number of challenges due to its complexity, especially in nanoparticles\cite{SAPRB}. Due to symmetry thin films present a special case for surface anisotropy, where the behavior is much simplified. Thin films have attracted a great deal of research interest over the past 50 years and so a large body of experimental data exists\cite{Chuang1994, Baek2003, Allenspach1992}. Nevertheless, achieving good experimental data on the temperature dependence of surface anisotropy requires the creation of very thin films with very sharp interfaces, which has only been technologically feasible within the last decade. This is because the influence of surface anisotropy is usually determined by varying the thickness of the magnetic layer, so that volume and surface contributions can be separated. For thick films the volume component strongly dominates the overall anisotropy, leading to a large degree of uncertainty in the strength of the surface contribution. Another problem arises with temperature dependent atomic migration, structural changes and interface mixing, which cause a change in the surface properties \cite{Dinia1998}.

\subsection*{Calculation of temperature dependent anisotropy}\label{sec:results}
The constrained Monte Carlo method allows for a thorough investigation of the temperature dependence of anisotropy in thin films with the N\'eel surface anisotropy. In the case of a perfect single crystal magnetic film with a face-centered-cubic (fcc) or simple cubic (sc) crystal structure with interfaces cut along the [001] direction, the on-site N\'eel surface anisotropy yields a purely uniaxial anisotropy. In order to simulate a section of such a thin film, a generic magnetic material with fcc ($T_{\mathrm{c}}=1300K$) or sc ($T_{\mathrm{c}}=1000K$) crystal structure was chosen. In order to eliminate edge effects within the film, periodic boundary conditions in the film plane were used. The surface anisotropy is normally found to be much stronger than bulk-type anisotropy, and so a value of $K_{\mathrm{s}} = 10 K_{\mathrm{u}}$ was chosen. When studying thin films with surface anisotropy, a number of basic combinations of anisotropies are possible. Principally, in the case of bulk uniaxial anisotropy, the surface and bulk anisotropies can have aligned or opposing easy axes, depending on the sign of the constant. Alternatively, a material could possess a cubic bulk anisotropy and uniaxial surface anisotropy. In the following we present calculations of temperature dependent effects in these thin film systems with surface anisotropy.

\begin{figure}[htpbc]
   \begin{center}
   \includegraphics[totalheight=0.25\textheight]{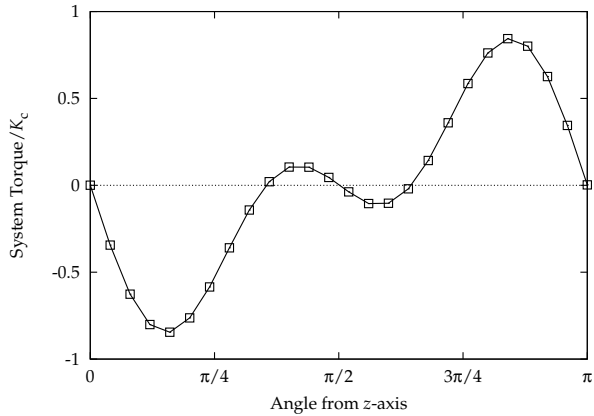}\\
   \caption{
	The torque curve  in a thin film of $L_{\mathrm{z}}=10$ atomic planes with fcc structure, cubic bulk anisotropy and N\'eel surface anisotropy perpendicular to the thin film plane and parallel to one of the bulk anisotropy easy axes for $T=10K$. The line provides a guide to the eye.
   }
   \label{fig:cub_uni}
   \end{center}
\end{figure}

\begin{figure}[htpbc]
   \begin{center}
   \includegraphics[totalheight=0.25\textheight]{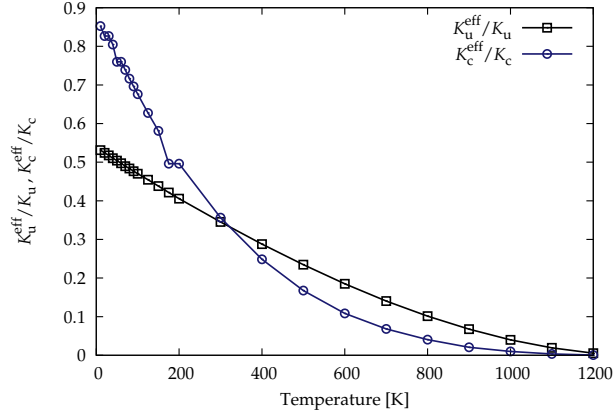}\\
   \caption{
	Temperature dependence of uniaxial and cubic effective anisotropies in a thin film
of $L_{\mathrm{z}}=10$ atomic planes with fcc structure, cubic bulk anisotropy and N\'eel surface anisotropy perpendicular to the thin film plane and parallel to one of the bulk anisotropy easy axes. The lines provide a guide to the eye. (Color Online)
   }
   \label{fig:Mixed}
   \end{center}
\end{figure}

Where both bulk and surface anisotropies are uniaxial, the torque curves are similar to ones presented in Fig.~\ref{fig:torque}, showing a  $\sin (2\theta)$ angular dependence.
In Fig.~\ref{fig:cub_uni} we present a more complicated situation, where the thin film has cubic bulk anisotropy and N\'eel surface anisotropy. One of the easy axes of the cubic anisotropy coincides with the surface anisotropy easy axis (perpendicular to the thin film surface). The torque curve clearly shows a summation of uniaxial and cubic anisotropy contributions. The temperature dependence of both contributions is presented in Fig.~\ref{fig:Mixed}. Similar to our results in the previous section, the cubic anisotropy exhibits a much stronger temperature dependence than the uniaxial (surface) part. Consequently, at low temperature the cubic anisotropy dominates while at high temperature the uniaxial surface anisotropy dominates.

\subsection*{Modeling of the reorientation transition in thin films.}
\begin{figure}[htpbc]
   \begin{center}
   \includegraphics[totalheight=0.25\textheight]{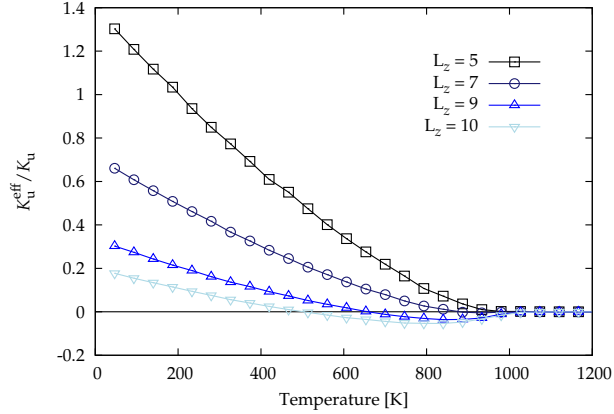}\\
   \caption{
	Temperature dependence of the effective anisotropy in thin films with in-plane bulk anisotropy and perpendicular to the plane surface anisotropy for various thin film thicknesses. The lines provide a guide to the eye. (Color Online)
   }
   \label{fig:Reorent}
   \end{center}
\end{figure}

The temperature dependence of the surface anisotropy  leads to a number of interesting effects, such as a temperature dependent reorientation of the magnetization direction from out-of-plane to in-plane and vice versa \cite{Schultz, Berger, Baek2003, Bruno1989, Allenspach1992, Enders, Dinia1998,Farle1998}.
Such an effect can occur when the easy directions of the surface and bulk anisotropies compete. At low temperatures the magnetization lies along the surface easy direction, e.g. perpendicular to the plane. As the temperature is increased the surface contribution to the anisotropy energy rapidly decreases, so the system magnetization lies along the bulk easy direction, e.g. in the plane. The temperature dependence of the effective anisotropy is plotted in Fig.~\ref{fig:Reorent} for different thin film thicknesses. Given the large difference in the surface and bulk anisotropy constants, the ultrathin films fail to show any reorientation transition. As the film thickness is increased the reorientation transition becomes more pronounced and occurs at a lower temperature. These results are comparable to mean-field calculations by Hucht \emph{et al} \cite{Hucht1997}.

\begin{figure}[htpbc]
	\begin{center}
	\includegraphics[totalheight=0.25\textheight]{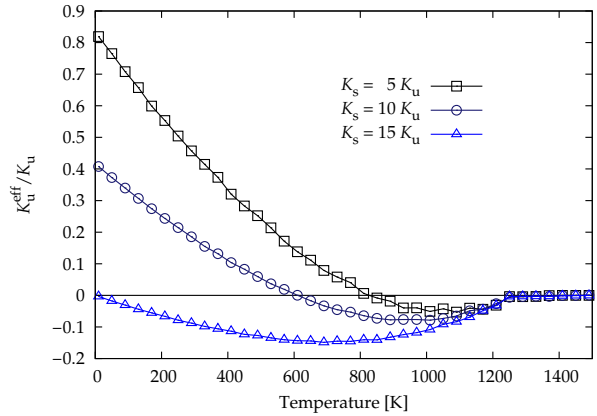}\\
	\caption[]
	{
		Temperature dependence of effective anisotropy for a thin film with competing surface (easy plane) and bulk  (parallel to $z$ axis) anisotropies for a system of $26 \times 26 \times 6$ unit cells for different magnitudes of $K_{\mathrm{s}}$. The lines provide a guide to the eye. (Color Online)
	}
	\label{fig:cmc_thin_film}
	\end{center}
\end{figure}

One other interesting property of the temperature dependent reorientation transition is that, depending on the choice of non-magnetic interface material, the temperature of the transition can be tuned. To illustrate this phenomenon, Fig.~\ref{fig:cmc_thin_film} shows a plot of the temperature dependence of the total anisotropy for different N\'eel anisotropy constants, emulating the effect of changing the interface material. Here the bulk anisotropy is assumed to be perpendicular to the plane, and the surface anisotropy is easy plane.

\subsection*{Temperature dependence of surface anisotropy}
The temperature dependent effective anisotropy in thin films with surface anisotropy is not an intrinsic parameter since it is strongly dependent on the thin film thickness. In this section we present two methods which enable the separation of the surface and bulk anisotropy contributions as a function of temperature in thin films, for simplicity, with parallel surface and bulk uniaxial anisotropy axes perpendicular to the thin film plane. This allows the extraction of the intrinsic uniaxial and surface contributions, independent of the thin film thickness.

\begin{figure}[htpbc]
   \begin{center}
   \includegraphics[totalheight=0.25\textheight]{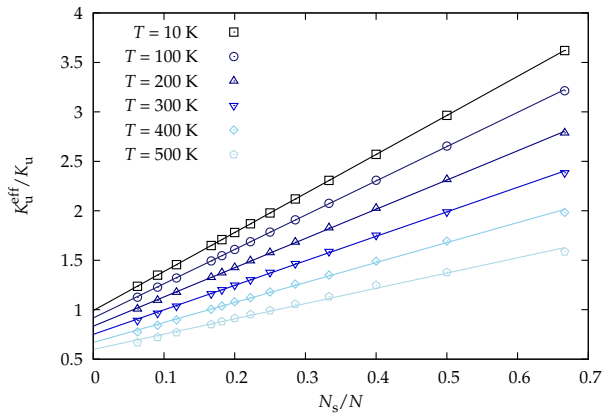}\\
   \caption{Scaling of the effective  anisotropy  with the ratio between surface and total number of atoms $N_{\mathrm{s}}/N$ in a thin film with sc structure, uniaxial bulk anisotropy and N\'eel surface anisotropy both perpendicular to the thin film plane. The lines show the fits to the data. (Color Online)}
   \label{fig:K_eff}
   \end{center}
\end{figure}

\begin{figure}[htpbc]
   \begin{center}
   \includegraphics[totalheight=0.25\textheight]{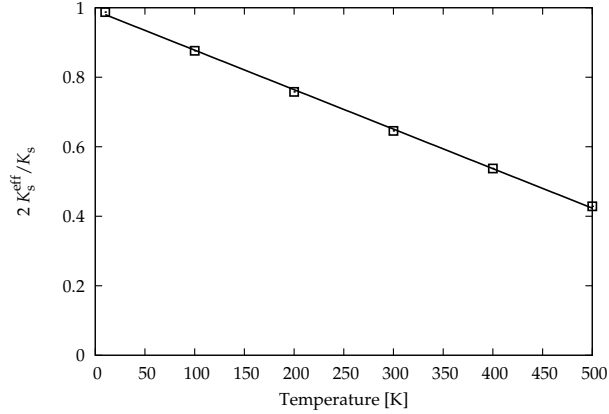}\\
   \caption{
	Temperature dependence of the effective surface anisotropy in a thin film with sc structure, uniaxial bulk anisotropy and N\'eel surface anisotropy perpendicular to the thin film plane determined from the size scaling of the effective anisotropy. The line is a linear fit to the data.
   }
   \label{fig:KsurfT}
   \end{center}
\end{figure}

The first method is based on the variation of the effective anisotropy with the number of surface atoms via the following well known expression\cite{Chappert}:
\begin{equation}
\label{scaling}
K_{\mathrm{eff}}=K_{\mathrm{u}}^{\mathrm{eff}}+\frac{N_{\mathrm{s}}}{N}(K_{\mathrm{s}}^{\mathrm{eff}}-K_{\mathrm{u}}^{\mathrm{eff}})
\end{equation}
where $N_{\mathrm{s}}$ and $N$ are the number of surface and total atoms, and $K_{\mathrm{s}}^{\mathrm{eff}}$ and $K_{\mathrm{u}}^{\mathrm{eff}}$ are the effective surface and bulk anisotropies, respectively. In Fig.\ref{fig:K_eff} we present results at different temperatures in thin films with a simple cubic lattice and different thicknesses. The data are perfectly scaled with the ratio $N_{\mathrm{s}}/N$ and for increasing film thickness the effective anisotropy tends towards the temperature dependent bulk value.  The fitting of the data to Eq.~(\ref{scaling}) allows the extraction of the surface anisotropy constant as a function of temperature, as presented in Fig.~\ref{fig:KsurfT}. In this system the surface anisotropy shows a linear decrease with temperature, similar to the experimental results \cite{Enders, Andre1995, Farle1997}.

\begin{figure}[htpbc]
   \begin{center}
   \includegraphics[totalheight=0.25\textheight] {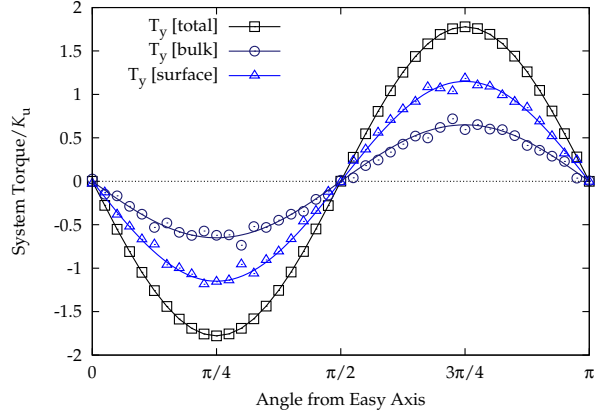}\\
   \caption{
	Simulated restoring torque in a thin film of $L_z=10$ atomic planes with sc structure, uniaxial bulk anisotropy and N\'eel surface anisotropy perpendicular to the thin film plane for $T=10 K$. The lines show a fit of the data to $\sin 2\theta$. (Color Online)
   }
   \label{fig:LD_sc_ez_TSC}
   \end{center}
\end{figure}

Since the surface atoms must be identified in order to calculate the N\'eel surface anisotropy, one can also resolve the surface and bulk contributions to the restoring torque curves. Fig.~\ref{fig:LD_sc_ez_TSC} shows the  torque curves for the total, bulk and surface parts, evaluated for the same thin film as above. As can be seen, the surface torque also follows the $\sin (2\theta)$ behavior and the effective surface anisotropy can therefore be extracted. The values of the surface anisotropy obtained through this method are comparable with the ones obtained through the scaling method shown in Fig.~\ref{fig:KsurfT}.

In practice, the surface torque is a noisy quantity due to the relatively small number of atoms. In order to obtain a good thermodynamic average it is generally necessary to use a large number of steps. However, the total torque converges much more rapidly and requires relatively few steps.

\subsection*{The scaling behavior of the anisotropy with magnetization}
The separation of surface and bulk contributions to the anisotropy allows the investigation of the temperature scaling of the surface anisotropy separately with respect to the surface magnetization, $M_{\mathrm{surface}}$, and bulk magnetization, $M_{\mathrm{bulk}}$. We should bear in mind that
the magnetization fluctuations on the surface are dependent on the bulk so that in general the corresponding scaling exponent is strongly system size dependent.

For an isolated surface layer the scaling of the surface anisotropy with respect to the surface magnetization should follow $K_{\mathrm{s}}^{\mathrm{eff}} \sim M_{\mathrm{surface}}^3$, as was found in the bulk case. In principle this effect could be measured experimentally using a monolayer of magnetic material, though such a structure is generally unstable at anything other than cryogenic temperatures.

\begin{figure}[htpbc]
	\begin{center}
	\includegraphics[
		totalheight=0.25\textheight
	]{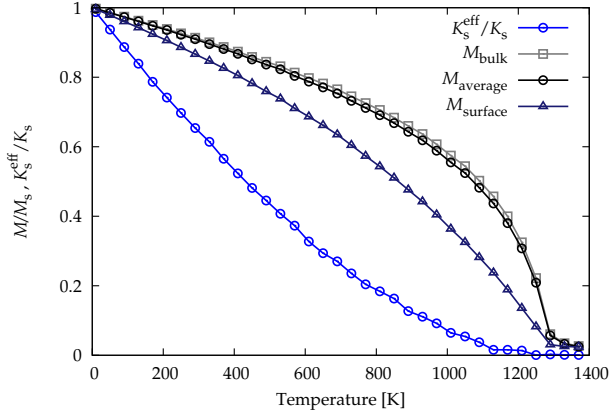}\\
	\caption
	{
		Plot of normalized magnetization and surface anisotropy against temperature for a  thin film with zero bulk anisotropy. The surface magnetization shows stronger criticality than the bulk and average magnetization. The lines provide a guide to the eye. (Color Online)
	}
	\label{fig:TFMvsT}
	\end{center}
\end{figure}

We have found that a magnetic thin film with zero bulk anisotropy also follows the $K_{\mathrm{s}}^{\mathrm{eff}} \sim M_{\mathrm{surface}}^3$ law, at least for the thin film thicknesses for which our calculations were feasible.  We have simulated a thin film system with fcc crystal structure, surface anisotropy perpendicular to the thin film plane and with zero bulk anisotropy.  This essentially ensures that the only anisotropic contribution to the Hamiltonian comes from the surface. The normalized  magnetization and surface anisotropy calculated via the torque method as a function of temperature are plotted in Fig.~\ref{fig:TFMvsT} for the system dimensions  $32 \times 32 \times 12$ unit cells. The surface, bulk and volume average magnetization are plotted, each having the same Curie temperature but with a different criticality, as previously reported by Binder et al \cite{Binder1974}.

\begin{figure}[htpbc]
	\begin{center}
	\includegraphics[
		totalheight=0.25\textheight]
	{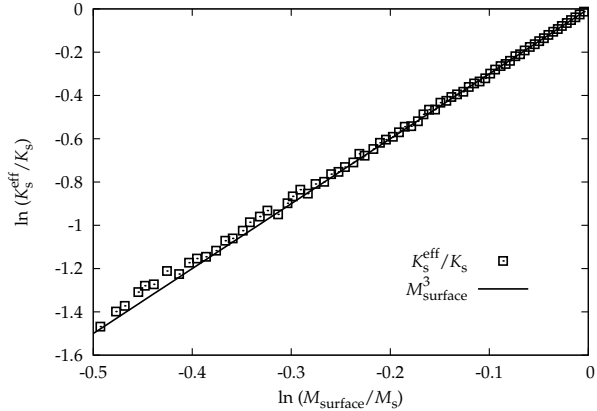}\\
	\caption
	{
		Plot of temperature scaling of surface anisotropy with surface magnetization, for a system with zero bulk anisotropy.
	}
	\label{fig:C5SMscaling}
	\end{center}
\end{figure}

The reduced criticality in the surface magnetization arises from a reduction in coordination number. An isolated surface layer would also have a reduced Curie temperature, but in our case  the surface layer is polarized by the bulk and thus has the same $T_{\mathrm{c}}$  of around 1300 K. Fig.~\ref{fig:C5SMscaling} shows the temperature scaling of the surface anisotropy with the surface magnetization showing a low temperature exponent of $K_{\mathrm{s}}^{\mathrm{eff}} \sim M_{\mathrm{surface}}^3$,
in excellent agreement with the Callen-Callen theory for single ion uniaxial anisotropy.

\begin{figure}[htpbc]
	\begin{center}
	\includegraphics[
		totalheight=0.25\textheight]
	{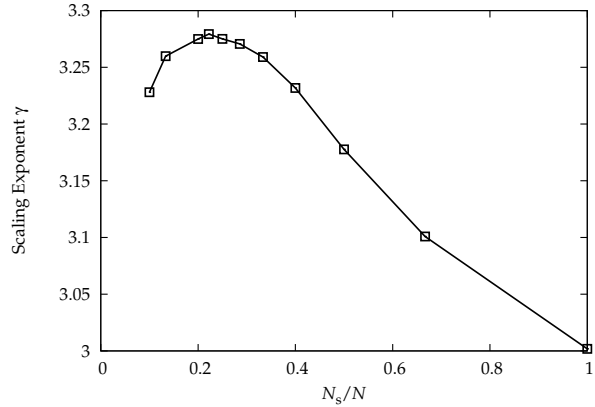}\\
	\caption
	{
		Plot of scaling exponent with thin film thickness in a thin film with $K_{\mathrm{s}}=10K_{\mathrm{u}}$, parallel out-of-plane easy axes and sc lattice. The line provides a guide to the eye.
	}
	\label{fig:SMTscaling}
	\end{center}
\end{figure}

The scaling of the total effective anisotropy in the presence of the
N\'eel surface anisotropy with total magnetization is unknown a-priori, and is coordination number, thin film thickness and material dependent. Nevertheless, it is this scaling which would be measured experimentally.
To illustrate this effect we have  calculated the effective scaling exponent $K_{\mathrm{u}}^{\mathrm{eff}} \sim M_{\mathrm{average}}^{\gamma}$ for a system with both uniaxial and surface anisotropies for different film thicknesses, as shown in Fig.~\ref{fig:SMTscaling}. For an isolated surface layer $N_{\mathrm{s}}=N$ the critical exponent $\gamma=3$ is recovered. The critical exponent has a maximum and should tend again to $\gamma=3$ value  for very thick films.

\section{Conclusion}\label{sec:conclusion}
We have developed a constrained Monte Carlo method, by which we can compute thermodynamic properties of magnetic systems as a function of the magnetization direction. We have shown its novel capability to evaluate the temperature dependence of the magnetic anisotropy, which is an important quantity for technological applications in hard magnetic materials. The method has been utilized to compute the temperature dependence of bulk magnetic anisotropy and we have recovered numerically the analytic scaling law of Callen and Callen.

The importance of the method resides in its potential to calculate temperature-dependent effective anisotropies in complex materials. The present challenge for modeling of magnetic materials is the multiscale approach, where the ab-initio information is passed to a different scale with the aim to model larger sizes of material, using, for example micromagnetics. The CMC method provides a real possibility to link the quantum mechanical scale with micromagnetics, via the parameterization of a classical Heisenberg Hamiltonian \cite{KazantsevaPRB2008}. This Hamiltonian can be used to calculate temperature dependent equilibrium behavior. To demonstrate this, we have applied the method to the calculation of the effective anisotropy bulk FePt, using the model parameterized through ab-initio calculations, and have recovered the experimentally observed $K_{\mathrm{FePt}}^{\mathrm{eff}} \sim M^{2.1}$ temperature scaling. Moreover, we have also noted the appearance of a flattening of the free energy surface at the energy maximum, an effect not previously seen with less subtle methods.

We have then applied the method to a variety of thin films with surface anisotropy, investigating the size dependent behavior and temperature dependence of surface anisotropy. We have shown the capability of the method to simulate the temperature-dependent magnetization reorientation transition. We also have shown that the method enables the separation of the temperature dependent surface anisotropy as an intrinsic system parameter, independent of the thin film thickness. Our results demonstrate a  linear temperature dependence of the surface anisotropy, consistent with the experimental results in Gd\cite{Andre1995, Farle1997},  Ni\cite{Farle1997} and Fe\cite{Enders} grown on different substrates. However, when comparing with our results, various factors should be taken into account, including structural changes with increased temperatures or a lattice mismatch which could influence the bulk anisotropy. One other possibility is that of an enhanced exchange interaction at the surface of the material \cite{Buruzs2007}. An increased exchange interaction would lead to an \emph{increase} in the criticality of the surface layer and commensurate reduction in the temperature dependence of the surface anisotropy. In the future, more complicated models could take into account these effects.

Finally, we have investigated the scaling of the anisotropy in thin films with magnetization. In thin films with zero bulk anisotropy, the surface anisotropy scales with the surface magnetization following the Callen-Callen law. In other cases we report no universal scaling behavior of the surface anisotropy.

To summarize, the constrained Monte Carlo method is a powerful tool, allowing to include both thermodynamic fluctuations and entropy into the evaluation of macroscopic quantities such as temperature dependent magnetic anisotropy. In the future we plan to apply the method to model more realistic systems within a general trend to large-scale material modeling aiming the design of novel materials with potential applications.

\section{Acknowledgements}\label{sec:acknowledgements}
This work has been supported by the grants MAT2007-66719-C03-01 and CS2008-023 from the Spanish Ministry of Science and Education, and EC FP7 [Grants No. NMP3-SL-2008-214469 (UltraMagnetron) and No. 214810 (FANTOMAS)]. The financial support from the European COST P-19 Action and from Seagate Technology, USA, is also gratefully acknowledged.

\appendix
\section{Jacobians in the trial moves}\label{sec:CMCderiv}

We are displacing two spins, whose combined
volume element is $d^2\sh_i\thinspace d^2\sh_j$.
However, we plan to readjust $\bsh_j$ to keep $\bMh$ constant.
Accordingly, we eliminate the variable $\bsh_j$ in favor of $\bMh$
and rewrite the volume element as
$\abs{J(\bsh_i,\bMh)}\thinspace d^2\sh_i\thinspace d^2\Mh$,
expressing $\bsh_j$ in terms of $\bMh$ and expecting the Jacobian
$J(\bsh_i,\bMh)$ to be nontrivial.
This allows us to view the trial move as taking place in the
$\bsh_i$, $\bMh$ variables.
The Boltzmann probability density in these variables is
$\rho \propto \abs{J}\exp(-\beta\HH)$.
Since our trial moves are reversible in $\bigl(\bsh_i,\bMh\bigr)$,
the Metropolis-Hastings test ratio \cite{Hastings1970} is just
$\rho'/\rho=$\mbox{$\abs{J'/J}\exp\bigl(-\beta\Delta\HH\bigr)$}.
The fact that we always keep $\bMh'\equiv \bMh$ has no bearing on
the argument and we use the same ratio to decide
whether to accept the move from $(\bsh_i,\bMh)$ to
$(\bsh_i',\bMh)$.

The first result we need is the relationship between a spherical
surface element and its projection in the $XY$ plane.
We have
\begin{gather}\label{proj}
    d^2\sh_i = \frac{d\sh_{ix}\thinspace d\sh_{iy}}{\abs{\sh_{iz}}}\quad, \quad
    d^2\sh_j = \frac{d\sh_{jx}\thinspace d\sh_{jy}}{\abs{\sh_{jz}}}\\
    \label{Mproj}
    d^2\Mh = \frac{d\Mh_x\thinspace d\Mh_y}{\abs{\Mh_z}}
    = d\Mh_x\thinspace d\Mh_y \quad.
\end{gather}
Too see this, note that for any unit vector $\bsh$ the angle between
the tangent plane to the sphere at $\bsh$ and the $XY$ plane is simply
$\theta= \cos^{-1}\left(\abs{\sh_z}\right)$, therefore the projected area
$d\sh_x\thinspace d\sh_y$ is $\abs{\cos(\theta)}\thinspace d^2\sh$
$=\abs{\sh_z}\thinspace d^2\sh$.
Equations \eqref{proj} and \eqref{Mproj} are the inverse
of this relation,
stated for the three vectors $\bsh_i$, $\bsh_j$, $\bMh$.
In the case of $\bMh$, Eq. \eqref{Mproj}, the denominator
$\abs{\Mh_z}$ is not necessary because we are locking $\bM$
in the positive $\bzh$ direction, therefore $\Mh_z\equiv 1$.

Next, we express $d\Mh_x$ and $d\Mh_y$ in terms of $d\sh_{ix}$,
$d\sh_{iy}$, $d\sh_{jx}$ and $d\sh_{jy}$.
We assume that $\bMh$ lies in the $\bzh$ direction
but we do not make the same assumption for $\bMh+d\bMh$.
For the {\em un}normalized vector $\bM$, we have
\begin{equation}\label{dM}
    dM_x= d\sh_{ix} + d\sh_{jx} \quad,\quad
    dM_y= d\sh_{iy} + d\sh_{jy}
\end{equation}
since the other spins are fixed.
%
%
For the normalized $\bMh$, we have
\begin{equation}\label{dMh}
    d\bMh = d\left(M^{-1}\bM\right) =
    M^{-1}\thinspace d\bM - M^{-2}\bM\thinspace dM
    \quad.
\end{equation}
We only need the $x$ and $y$ component of \eqref{dMh}.
Since $\bM$ lies in the $+\bzh$ direction, the last term
contributes nothing to the in-plane components
and we don't need to calculate $dM$ (although
it would be easy to do so).
The remaining term gives
\begin{equation}\label{dMxdMypre}
    d\Mh_x = M^{-1}\thinspace dM_x \quad,\quad
    d\Mh_y = M^{-1}\thinspace dM_y \quad.
\end{equation}
We substitute \eqref{dM} in \eqref{dMxdMypre} and replace $M$
by $M_z$:
\begin{equation}\label{dMxdMy}\begin{split}
    d\Mh_x &= M_z^{-1}\thinspace d\sh_{ix} + M_z^{-1}\thinspace d\sh_{jx}
    \\
    d\Mh_y &= M_z^{-1}\thinspace d\sh_{iy} + M_z^{-1}\thinspace d\sh_{jy}
    \quad.
\end{split}\end{equation}
We rewrite \eqref{dMxdMy} in matrix form,
\begin{equation}\label{d2Mmat}
    \begin{bmatrix}
	d\sh_{ix} \\ d\sh_{iy} \\ d\Mh_x \\ d\Mh_y
    \end{bmatrix}
    = \begin{bmatrix}
	1 & 0 & 0 & 0 \\
	0 & 1 & 0 & 0 \\
	M_z^{-1} & 0 & M_z^{-1} & 0 \\
	0 & M_z^{-1} & 0 & M_z^{-1}
    \end{bmatrix}
    \begin{bmatrix}
	d\sh_{ix} \\ d\sh_{iy} \\ d\sh_{jx} \\ d\sh_{jy}
    \end{bmatrix}
    \quad.
\end{equation}
The Jacobian of the change of variables is the determinant of the
matrix in \eqref{d2Mmat}, namely $M_z^{-2}$.
Therefore, the volume elements are related by
\begin{equation}\label{volback}
    d\sh_{ix}\thinspace d\sh_{iy} \thinspace
    d\Mh_x \thinspace d\Mh_y =
    M_z^{-2} \medspace
    d\sh_{ix}\thinspace d\sh_{iy} \thinspace
    d\sh_{jx} \thinspace d\sh_{jy}
\end{equation}
and the inverse relation is
\begin{equation}\label{volfwd}
    d\sh_{ix}\thinspace d\sh_{iy} \thinspace
    d\sh_{jx} \thinspace d\sh_{jy} =
    M_z^2 \medspace
    d\sh_{ix}\thinspace d\sh_{iy} \thinspace
    d\Mh_x \thinspace d\Mh_y
    \quad.
\end{equation}
Dividing by $\abs{\sh_{iz}}\thinspace\abs{\sh_{jz}}$ and using
\eqref{proj}, we obtain
\begin{equation}\label{vol}
    d^2\sh_i\thinspace d^2\sh_j = \frac{M_z^2}{\abs{\sh_{jz}}} \medspace
    d^2\sh_i \thinspace d^2\Mh
    \quad
\end{equation}
Thus the Jacobian is $\abs{J}= M_z^2/\abs{\sh_{jz}}$
and we accept each trial move with probability
\begin{equation}\label{Metropo}\begin{split}
    P&= \min\left[1, \abs{J'/J}\exp\bigl(-\beta\Delta\HH\bigr)\right] \\
	&= \min\Biggl[1,
	\biggl(\frac{M_z'}{M_z}\biggr)^2
	\frac{\abs{\sh_{jz}}}{\abs{\sh_{jz}'}}
	\exp\bigl(-\beta\Delta\HH\bigr)
    \Biggr]
\end{split}\end{equation}
as stated in step \ref{alg:Boltzmann} of the algorithm.

\section{Ergodicity}\label{sec:ergodic}

We show that every admissible state
(i.e. with total magnetization along $+\bzh$)
can be reached from any other admissible state by a sequence of constrained
Monte Carlo trial moves.
In this context we may ignore the Metropolis acceptance test
at step \ref{alg:accept}
of the algorithm, but the trial moves must still
pass the kinematic tests at steps \ref{alg:adjustz} and
\ref{alg:Mz}.
The proof relies on two lemmas:

\begin{lemma}
    Any spin with a negative $z$ component
    can be moved to the hemisphere $\sh_z> 0$
    by a sequence of trial moves,
    without any other spin crossing the plane $z=0$.
    \label{lemma:positive}
\end{lemma}

\begin{lemma}
    Assuming all the spins are in the hemisphere $\sh_z>0$,
    one of them can be moved to the $z$ axis by
    a sequence of trial moves, without any spin
    crossing the plane $z=0$.
    \label{lemma:alignz}
\end{lemma}

Repeated applications of Lemma \ref{lemma:positive} allow us
to move the spins one by one to the hemisphere $\sh_z>0$, at
which point Lemma \ref{lemma:alignz} becomes applicable.

Repeated applications of Lemma \ref{lemma:alignz} allow us
to move {\em all} the spins to the positive $z$ axis.
After the first application we have one spin along $+\bzh$;
the remaining $N-1$ spins are perturbed in the process, but they
remain in the upper hemisphere and their final magnetization
must also point along $+\bzh$.
That is, they form an admissible state of $N-1$ spins,
Lemma \ref{lemma:alignz} becomes applicable to them
and we can iterate.
Thus, any admissible state can be collapsed to
the saturated state by a sequence of trial moves.
By chaining such a collapsing sequence with the inverse of another one,
we can connect any two admissible states.

\subsection*{Proof of Lemma \ref{lemma:alignz}}
We use trial moves where the primary spin does not cross
the plane $z=0$.
We write $\bs_i$ (lower case) for the projection
$(\sh_{ix},\sh_{iy})$ in the $XY$ plane.
It is enough to consider the $\bs_i$ since the $z$ components are
uniquely determined by them, $\sh_{iz}=+\sqrt{1-\norm{\bs_i}^2}$.
The trial moves can not fail at step \ref{alg:Mz}, leaving
only step \ref{alg:adjustz} to consider.
A trial move then consists in displacing two points $\bs_i,\bs_j$
by equal and opposite amounts, while keeping both points in the
unit disk, $\norm{\bs_i}, \norm{\bs_j} \leq 1$.

We want to show that, for any set of $N$ points in the unit disk
with centroid $(0,0)$, one of the points can be shifted to $(0,0)$ by
a combination of such moves.
In fact we have a slightly more general result:
in any set of $N$ points, one can be moved to the centroid,
whether or not the centroid is $(0,0)$.

\begin{figure} [htpbc]
	\begin{center}
	\includegraphics[
		width=0.40\textwidth,
	]{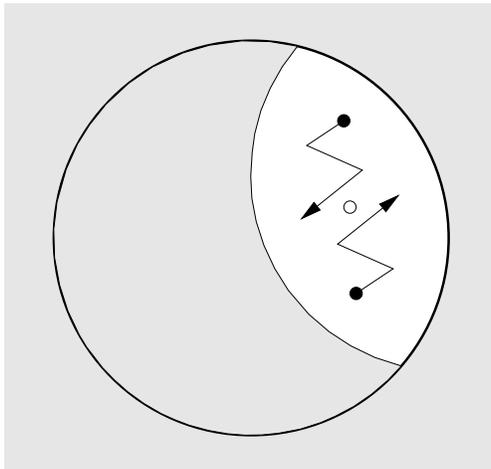}\\
	\caption[
		Admissible zone for projected trial moves in Lemma
		\ref{lemma:alignz}
	anisotropy
	]{
		Admissible zone for projected trial moves in
		the proof of Lemma \ref{lemma:alignz}.
		The projections of the two spins (black circles) can move
		symmetrically about their centroid (white circle)
		within the region shown in white.
	}
	\label{fig:lens}
	\end{center}
\end{figure}

We proceed by induction on $N$.
For $N<2$ there is nothing to prove.
For $N=2$, let $\bc=(\bs_1+\bs_2)/2$ be the centroid.
Since $\bc$ is invariant, the condition $\norm{\bs_2}\leq 1$ can be
expressed in terms of $\bs_1$, $\norm{2\bc-\bs_1}\leq 1$.
Similarly we have $\norm{2\bc-\bs_2}\leq 1$.
This means that we can move the two points symmetrically about $\bc$,
as long as we keep them both within
a lens-shaped ``admissible zone'', which is the intersection of
the unit disk and of its inversion through $\bc$.
This is shown in Fig.~\ref{fig:lens}.
The admissible zone, being convex, contains $\bc$.
Therefore we can move both points to $\bc$, using several sub-moves if necessary.

For $N>2$ we assume by the induction hypothesis that the
centroid $\bc_{N-1}$ of the first $N-1$ points is already
occupied by one of the $\bs_k$.
The full centroid $\bc_N=(\bs_N+(N-1)\bs_k)/N$ lies on the
segment $[\bs_k,\bs_N]$ and is in the admissible zone
of $\bs_k, \bs_N$.
Therefore we can move either point to the centroid.
This proves the lemma.

\subsection*{Proof of Lemma \ref{lemma:positive}}
We want to reduce by one the number of spins with $\sh_z\leq 0$.
Our rules allow trial moves where the primary spin
flips its $z$ component and the secondary spin does nothing,
but we would like to avoid these potentially low probability moves.
Our strategy is to move the offending spin close to the plane $z=0$
so as to require only a small jump in $\sh_z$ in the final move.
We proceed as in the proof of Lemma \ref{lemma:alignz}, using
projections in the $XY$ plane, but this time we have to avoid
rejection at step \ref{alg:Mz} of the algorithm.

Consider an admissible state where spin $i$ has $\sh_{iz}<0$.
Choose a $j$ such that $\sh_{jz}>0$ (there must be one,
since $M_z$ is positive).
We use trial moves with $\bsh_i$ as the primary spin
and $\bsh_j$ as the compensation spin.
Let $\bs_i, \bs_j$ be the projections in the $XY$ plane.
The two points $\bs_i,\bs_j$ are restricted to a lens-shaped
admissible region as in the proof of Lemma \ref{lemma:alignz},
but there may be a further restriction to ensure that
$M_z$ remains positive.

\begin{figure} [htpbc]
	\begin{center}
	\includegraphics[
		width=0.40\textwidth,
	]{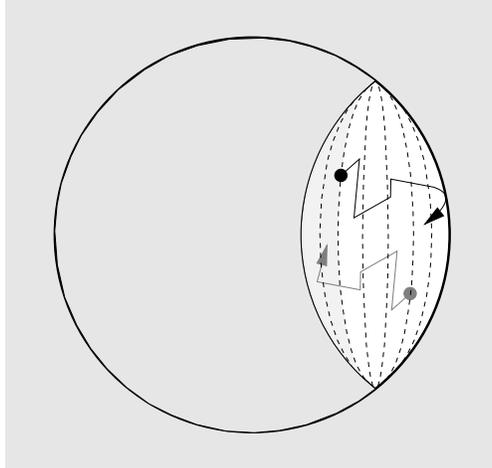}\\
	\caption[
		Admissible zone for projected trial moves in
		Lemma \ref{lemma:positive}
	]{
		Admissible zone for projected trial moves
		in the proof of Lemma \ref{lemma:positive}.
		The dashed lines are contours of
		$\sh_{iz}+\sh_{jz}$ as a function
		of $\sh_{ix},\sh_{iy}$.
		A possible path for the projection of the
		primary spin is shown in black, with
		the matching path of the secondary spin
		in grey.
		If the primary spin stays in the white region the total
		magnetization $M_z$ cannot become negative.
	}
	\label{fig:lens2}
	\end{center}
\end{figure}

As before, we express $\bs_j$ in terms of $\bs_i$,
$\bs_j= 2\bc-\bs_i$.
The contribution of spins $i$ and $j$ to $M_z$ is then
\begin{equation}\label{eq:ijMz}
    \sh_{iz}+\sh_{jz} = -\sqrt{1-\norm{\bs_i}^2} +
    \sqrt{1-\norm{2\bc-\bs_i}^2} \text{ .}
\end{equation}
It can be shown that the contours of \eqref{eq:ijMz}
in the $(s_{ix},s_{iy})$ plane are arcs of ellipses centered
on $\bc$ and tangent to the admissible zone, as shown in
Fig.~\ref{fig:lens2},
and that \eqref{eq:ijMz} increases monotonically
as $\bs_i$ moves from the interior edge to the exterior edge
of the admissible zone.
If $\bs_i$ stays on the distal side of its starting contour, the value
of $M_z$ cannot become negative.
This is represented by the white region in Fig.~\ref{fig:lens2}.
We see that there is an ample supply of paths that take
$\bs_i$ arbitrarily close to the edge of the unit disk.
At that point a final trial move can take $\bsh_i$ across
the plane $\sh_{iz}=0$.
The only other spin involved, $\bsh_j$, remains in the
positive $z$ hemisphere throughout.
This proves Lemma \ref{lemma:positive}.

\section{Macroscopic Torques}\label{cmc:torques}

The magnetization direction $\bMh$ plays the role of a macroscopic
parameter.
In this section we identify the generalized forces acting on
$\bMh$ (torques) with the derivatives of the Helmholtz free energy.

By way of introduction, consider a thermodynamic system
with an external macroscopic parameter $q$
and microscopic states $\xi_\alpha$.
We treat $\alpha$ as a discrete index for simplicity.
The energy of each microscopic state is a function of $q$;
the partition function and the Helmholtz free energy are \cite{Goodstein1975}
\begin{equation}\label{eq:Zpartition}
	\mathcal{Z}(q) = \sum_\alpha \exp\bigl(-\beta \HH(\xi_\alpha, q)\bigr)
	\thickspace,
\end{equation}
\begin{equation}\label{eq:Helmholtz}
	\mathcal{F}(q) = -\beta^{-1}\Ln\bigl(\mathcal{Z}(q)\bigr)
	\thickspace.
\end{equation}
Differentiation of \eqref{eq:Helmholtz} with respect to $q$
yields immediately
\begin{equation}\label{eq:genforce}\begin{split}
	-d\mathcal{F}/dq &= \sum_\alpha
	\mathcal{Z}^{-1} \exp\bigl(-\beta \HH(\xi_\alpha, q)\bigr)
	\medspace \frac{-\partial \HH(\xi_\alpha,q)}{\partial q} \\
	&= \bigl\langle -\partial \HH/\partial q \bigr\rangle
\end{split}\end{equation}
where in the second line of \eqref{eq:genforce} we recognized the
sum over microstates as a thermodynamic average, since the
weights $\mathcal{Z}^{-1}\exp\bigl(-\beta \HH(\xi_\alpha, q)\bigr)$
are the Boltzmann probabilities.
Thus the mean force conjugate to $q$ is the negative derivative with
respect to $q$ of the Helmholtz free energy $\mathcal{F}(q)$.

The right-hand side of \eqref{eq:genforce}, being a thermodynamic
average, can be computed by the Metropolis algorithm.
As a general rule, derivatives and differences of free energies
are computable in this way, even though the free energy
itself is not.

The argument is not directly applicable to our case since
$\bMh$ is not a parameter in the Hamiltonian,
but a restriction on the set of admissible states.
One could in principle set up a system of $2N-2$ coordinates
that forms a complete set with $\bMh$
and treat $\bMh$ as a parameter,
but we never did this.
Instead, we picked a temporary coordinate system for each
Monte Carlo move, to be discarded after the move.
Indeed, the whole of appendix \ref{sec:CMCderiv}
is a stratagem to avoid setting up global coordinates.

However we can recover a result similar to \eqref{eq:genforce}.
Given any two directions $\bMh$ and $\bMh'$ we can
find a rotation $\RR$ that sends $\bMh$ to $\bMh'$.
If we apply this rotation globally to every spin $\bsh_i$ that constitutes
a microstate $\xi$, we obtain a measure-preserving
bijection between the two admissible manifolds.
This allows us to replace sums over the microstates
of the $\bMh'$ manifold by sums over the microstates (rotated) of
the $\bMh$ manifold.
In particular, the partition function
$\mathcal{Z}'=\mathcal{Z}(\bMh')$ is
\begin{equation}\label{eq:zprime1}
	\mathcal{Z}' = \sum_\alpha
		\exp\bigl(-\beta \HH(\RR\xi_\alpha)\bigr)
	\thickspace.
\end{equation}
Dividing by $\mathcal{Z}$ and factoring out
an $\exp\bigl(-\beta \HH(\xi_\alpha)\bigr)$ from each term,
\begin{multline}\label{eq:zprime2}
	\mathcal{Z}'/\mathcal{Z} = \sum_\alpha
		\mathcal{Z}^{-1} \exp\bigl(-\beta \HH(\xi_\alpha)\bigl)\times\\
		\exp\bigl[-\beta\bigl(
			\HH(\RR\xi_\alpha)-\HH(\xi_\alpha)
		\bigr)\bigr]
	\thickspace,
\end{multline}
or
\begin{equation}\label{eq:zprime3}
	\mathcal{Z}'/\mathcal{Z} = \Bigl\langle \exp\bigl[-\beta\bigl(
		\HH(\RR\xi)-\HH(\xi)
	\bigr)\bigr] \Bigr\rangle
	\thickspace,
\end{equation}
where the sums in (\ref{eq:zprime1}--\ref{eq:zprime2})
and the thermodynamic average in \eqref{eq:zprime3} are
over the admissible manifold of $\bMh$.
Taking logarithms and dividing by $-\beta$, we have
\begin{equation}\label{eq:Fprime}
	\mathcal{F}' - \mathcal{F} =
	- \beta^{-1}\Ln\Bigl[ \Bigl\langle
		\exp\bigl[-\beta\bigl(
			\HH(\RR\xi) - \HH(\xi)
		\bigr)\bigr]
	\Bigl\rangle \Bigr]
\end{equation}
where again the thermodynamic average is over the manifold of $\bMh$.

Specializing to an infinitesimal rotation,
$\RR \bv \equiv \bv + d\theta \bnh\times\bv + O(d\theta^2)$ and
writing each microstate $\xi$ as a collection of spins $\bsh_i$,
the energy difference in the argument of the exponential is
\begin{align}
	\HH(\RR\xi)-\HH(\xi) &= d\theta \sum_i
		\bigl(\bnh\times\bsh_i\bigr)\cdot
		\frac{\partial \HH}{\partial \bsh_i}
		+ O(d\theta^2) \\
	&= d\theta\bnh\cdot \sum_i \biggl(
		\bsh_i\times\frac{\partial \HH}{\partial \bsh_i}
	\biggr) + O(d\theta^2)
	\thickspace.
\end{align}
As $d\theta$ approches zero the logarithm and exponential in \eqref{eq:Fprime}
become a no-op,
\begin{equation}\label{eq:FprimeR}
	\mathcal{F}'-\mathcal{F} = d\theta\bnh\cdot\biggl\langle
		\sum_i\bsh_i\times \frac{\partial \HH}{\partial \bsh_i}
	\biggr\rangle + O(d\theta^2)
	\thickspace.
\end{equation}
Meanwhile, $\bMh'$ is related to $\bMh$ by the same infinitesimal
rotation and we can expand $\mathcal{F}'=\mathcal{F}(\bMh')$ in
a Taylor series.
\begin{align}\label{eq:FTaylor}
    \mathcal{F}' &= \mathcal{F}\bigl(
	\bMh + d\theta\bnh\times\bMh + O(d\theta^2)
    \bigr) \\
    &= \mathcal{F} +
	d\theta\bigl(\bnh\times\bMh\bigr)\cdot
	\frac{\partial \mathcal{F}}{\partial \bMh}
    + O(d\theta^2) \\
    \label{eq:FprimeL}
    &= \mathcal{F} + d\theta\bnh\cdot\biggl(
	\bMh\times\frac{\partial \mathcal{F}}{\partial \bMh}
    \biggr) + O(d\theta^2)
\end{align}
Combining the terms of order $d\theta$ in \eqref{eq:FprimeR}
and \eqref{eq:FprimeL}, we obtain
\begin{equation}
    \bnh\cdot\biggl\langle
	\sum_i \bsh_i\times \frac{\partial \HH}{\partial \bsh_i}
    \biggr\rangle =
    \bnh\cdot\biggl(
	\bMh\times\frac{\partial \mathcal{F}}{\partial \bMh}
    \biggr)
    \medspace .
\end{equation}
Since the equality holds for all $\bnh$, we have finally
\begin{equation}\label{eq:avgtorque}
	\biggl\langle
		\sum_i \bsh_i\times \frac{\partial \HH}{\partial \bsh_i}
	\biggl\rangle
	=
	\bMh\times\frac{\partial \mathcal{F}}{\partial \bMh}
    \medspace .
\end{equation}

\end{document}